\documentclass[mathleft
]{an}
\usepackage{graphicx}
\usepackage{times}
\begin{document}

\Pagespan{789}{}
\Yearpublication{2006}%
\Yearsubmission{2005}%
\Month{11}%
\Volume{999}%
\Issue{88}%

\title{Cosmology with AGN: can we use quasars as standard candles?}

\author{G. Risaliti\inst{1,2}\fnmsep\thanks{Corresponding author:
  \email{risaliti@arcetri.astro.it}\newline}
\and  E. Lusso\inst{2}
}
\titlerunning{Quasars as standard candles}
\authorrunning{G. Risaliti \& E. Lusso}
\institute{
Department of Physics and Astronomy, University of Florence, Via G. Sansone 1, 50019 Sesto Fiorentino (FI) , Italy
\and 
INAF - Arcetri Astrophysical Observatory, Largo E. Fermi 5, 50126 Firenze, Italy}

\received{30 May 2005}
\accepted{11 Nov 2005}
\publonline{later}


\abstract{
The non-linear relation between X-ray and UV luminosity in quasars can be used to estimate their distance. Recently, we have shown that despite the large dispersion of the relation, a Hubble Diagram made of large samples of quasars can provide unique constraints on cosmology at high redshift. Furthermore, the dispersion of the relation is heavily affected by measurement errors: until now we have used serendipitous X-ray observations, but dedicated observations would significantly increase the precision of the distance estimates. We discuss the future role of XMM in this new field, showing (1) the fundamental contribution of the Serendipitous Source Catalogue and of large surveys, and (2) the breakthrough advancements we may achieve with the observation of a large number of SDSS quasars at high redshift: every $\sim$10 quasars observed at z$\sim$3 would be equivalent to discovering a supernova at that redshift.} 
\maketitle

\section{Introduction}

The relation between $\alpha_{OX}$ (the slope of a power law connecting the ($\nu,f_\nu$) points at 2,500~\AA~ and 2~keV in the Spectral Energy Distribution (SED) of a quasar) and the luminosity $L_{UV}$ at 2500~\AA~ implies that optically more luminous quasars are relatively less luminous in the X-rays: for example, an increase by a factor of 10 in  L$_{UV}$ implies an increase of only a factor of $\sim$4 in the X-ray luminosity at 2~keV, $L_X$. An equivalent way to describe the same result is by a luminosity-luminosity relation: $\log(L_X)=\alpha\times\log(L_{UV})+\beta$, with $\alpha<1$. The recent works on the largest samples available for these analysis (Just et al.~2007 for a collection of more than 300 quasars over a wide range in z and luminosity; Young, Risaliti \& Elvis~2010 for $\sim500$ quasars with SDSS and XMM spectra, Lusso et al.~2010 for the COSMOS quasar sample) agree on a value of $\alpha\sim0.6$. The observed dispersion of the relation is of the order of 0.35~dex.

The non-linear relation between the X-ray and UV luminosities of quasars can be used as a  distance indicator, allowing the construction of  a Hubble diagram for quasars. However, its large scatter has deterred attempts to use this relation for cosmology. The situation has changed with our first work on this topic (Risaliti \& Lusso~2015, ApJ 815, 33, hereafter RL15) where we used $\sim$1,000 sources with literature data in UV and X-rays, cleaned from X-ray absorption and dust-reddening effects, and obtained several relevant results:\\
(a) the UV to X-ray relation, $\log(L_X)\sim8.5+0.6\times\log(L_{UV})$ does not show any dependence with redshift, as needed to use it as distance estimator (Fig.~1); \\
(b) the quasars Hubble diagram is in excellent agreement with that of SN1a in the common redshift range (z$<$1.4), and extends to higher redshifts in agreement with the standard $\Lambda$CDM model. It also provides a significantly improved measurement of $\Omega_M$ and $\Omega_\Lambda$ with pure distance indicators; \\
(c) Simulations show that future large X-ray all-sky surveys (such as those of eROSITA) will match the available optical/UV samples (such as the SDSS). With tens to hundreds thousand quasars, this will become an extremely precise probe of cosmological models, and, together with Gamma Ray Bursts (Ghirlanda et al.~2012, Demianski et al.~2016) the only one sensitive to the expansion rate of the Universe at redshifts 3-6.  \\
\begin{figure}
\includegraphics[width=83mm,height=60mm]{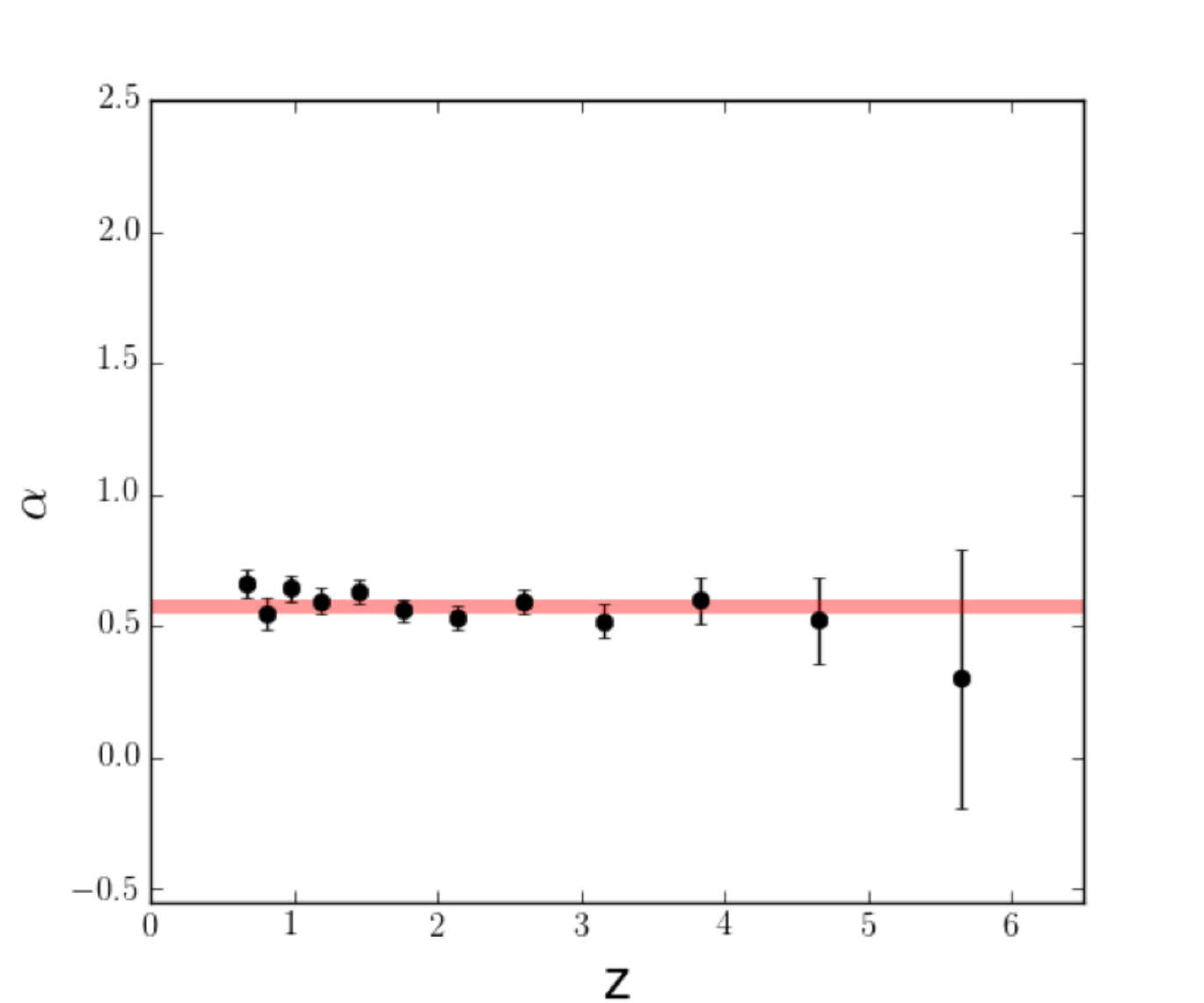}
\caption{Slope of the L$_X$-L$_{UV}$ relation calculated with subsamples in small redshift intervals. The horizontal line shows the mean value and its error. Within each interval the differences in distance are negligible with respect to the dispersion of the relation. Therefore, luminosities can be replaced by fluxes, and the measurements are independent from the cosmological parameters.}
\label{label1}
\end{figure}

\section{The dispersion of the L$_X$-L$_{UV}$ relation in quasars}

In the past months, we have extended our study by including new quasar samples obtained cross-correlating the Sloan Digital Sky Survey (SDSS)  quasar catalogues from Data Release 7 (DR7, Shen et al.~2010) and DR12 (DR12, Paris et al.~2016) with the 3XMM-DR6 catalogue of serendipitous XMM sources (Rosen et al.~2016), for a total of $\sim7,500$ new sources. This allowed us, for the first time, to work on a large and homogeneous sample, with enough statistics to apply strong quality cuts to our sources, and analyze only those with high S/N and reliable measurements both in X-rays and UV spectra. The results of this analysis have been rather surprising: most of the observed dispersion of the L$_X$-L$_{UV}$ relation is due to observational issues (Fig.~2), with the most important contribution from the uncertainty on the monochromatic flux at 2~keV, while the intrinsic dispersion is smaller than 0.20~dex. The details of this analysis are presented in Lusso \& Risaliti 2016, and Lusso et al.~2017 (in prep.) Here we just summarize the results, focusing on the implications for cosmological measurements. 
\begin{figure}
\includegraphics[width=83mm]{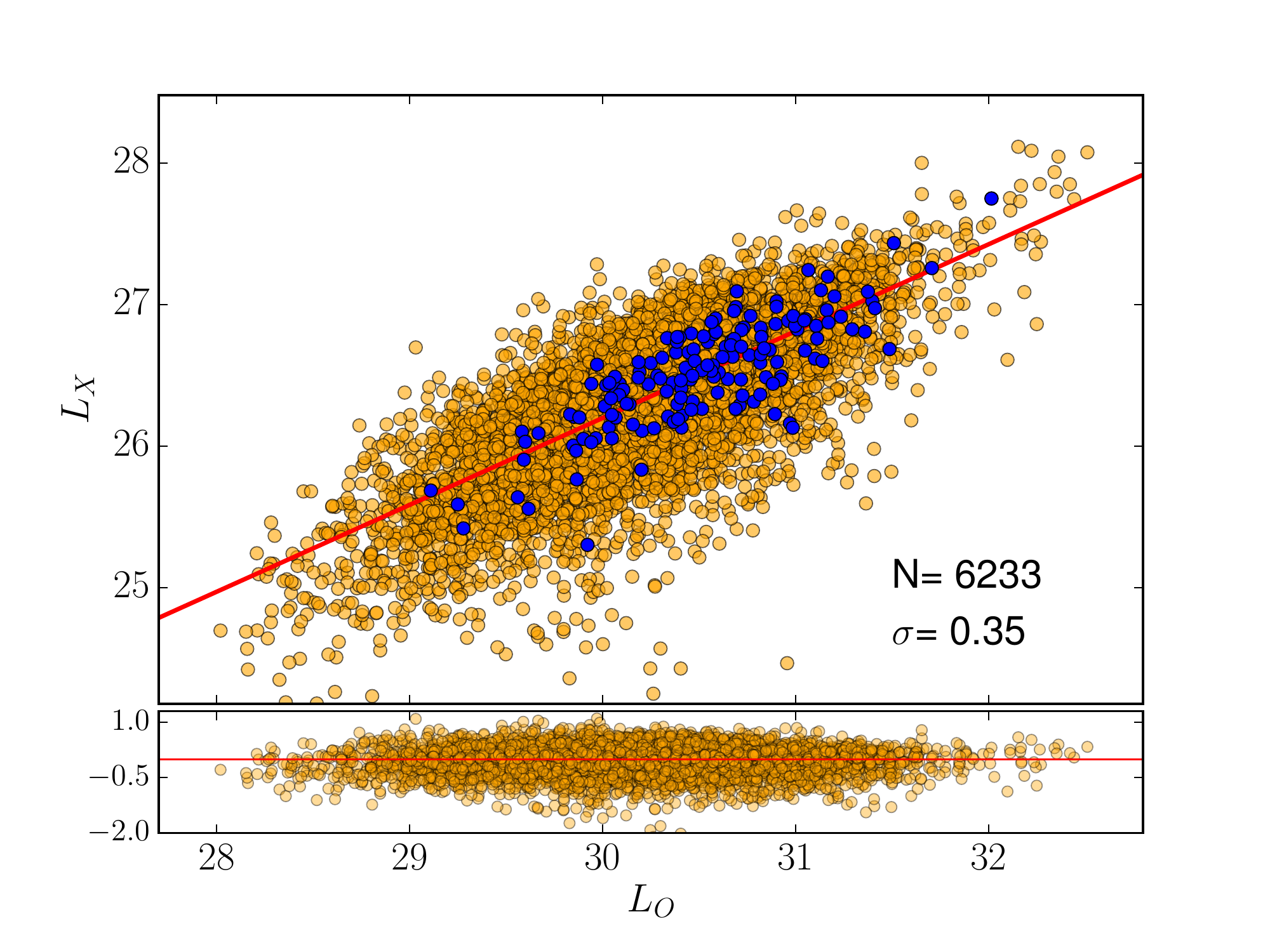}
\caption{L$_X$-L$_{UV}$ relation for a sample of more than 6,000 quasars with X-ray and UV measurements (gold points), and for a subsample of $\sim100$ objects with high-quality X-ray observations (blue points). The dispersion for this subsample os 0.21~dex.}
\label{label1}
\end{figure}

In order to understand the possible use of quasars as standard candles, it is important to distinguish between the {\em intrinsic} physical dispersion, $\sigma_{{\rm INT}}$, and the smallest dispersion we can reasonably observe, $\sigma_{{\rm OBS}}$. The former is an indication of how tight is the physical connection between the UV-emitting disk and the X-ray emitting corona. The latter is instead related to the smallest uncertainty on the distance we can obtain from UV and X-ray flux measurements of quasars. We selected a subsample of about 100 quasars with high S/N X-ray spectra, a redshift range between 0.5 and 2.7 (in order to directly measure the 2500~\AA~ flux from the SDSS spectrum, without extrapolations), no indication of absorption/extinction in X-rays and in optical/UV, and with two independent XMM observations. We fitted the L$_X$-L$_{UV}$ relation adopting a standard $\Lambda$CDM cosmology (at this stage, we are not interested in testing the cosmological model, but only to investigate the physical relation) using (a) the flux from the longest X-ray observation, and (b) the average from the two observations. The measured dispersions are 0.21~dex and 0.18~dex, respectively. The observed dispersion is due to a contribution from X-ray measurements, $\sigma_X$, and a residual one due to all the other (independent) effects, $\sigma_{{\rm RES}}$: $\sigma_{{\rm OBS}}=\sqrt{\sigma_X^2+\sigma_{{\rm RES}}^2}$. Averaging the two observations reduces $\sigma_X$ by a factor $\sqrt{2}$. The residual dispersion $\sigma_{{\rm RES}}$ can then be estimated as $\sim$0.14~dex. $\sigma_{{\rm RES}}$ includes the uncertainties on the 2500~\AA~ flux (which are in most cases as low as $\sim$0.05~dex, Kovlowski et al.~2010) and, most importantly, the spread due to the random inclination of the accretion disk with respect to the observer. Even considering a bias towards face-on systems in the optically selected, flux limited SDSS samples (Risaliti et al.~2010, Bisogni et al.~2016), the dispersion due to this effect is of the order of the residual dispersion $\sigma_{RES}$. From this analysis we then draw two important conclusions:\\
\begin{itemize}
\item We explained $\sim$all the observed dispersion with contributions from observational errors, variability, and disk inclination. This implies a very tight intrinsic L$_X$-L$_{UV}$, due to a universal physical process linking the accretion disk and the X-ray corona in quasars.
\item The smallest dispersion  we can reasonably observe is $\sigma_{{\rm OBS}}$=0.20-0.21 for single X-ray observations, and $\sigma_{{\rm OBS}}$=0.14 with several repeated observations. When the relation is used for cosmology, the same uncertainty is propagated to the distance estimates. This can be compared with the dispersion of the Hubble Diagram of supernovae 1A at z$\sim$1, $\sigma_{{\rm SN}}\sim 0.07$. We conclude that 9-10 quasars have the same amount of cosmological information as one SN1A. 
\end{itemize}
\section{The Hubble Diagram for quasars}
We merged all the literature samples discussed in Risaliti \& Lusso~2015 with the cross-match between SDSS quasar catalogues and the 3XMM-DR6 serendipitous source catalogue, obtaining a sample of almost 8,000 objects, after excluding radio-loud and Broad Absorption Line (BAL) quasars. We then rejected all the objects with a X-ray and/or UV absorption and those possible affected by an ``Eddington bias'' (i.e. objects with an average X-ray flux below our detection threshold, but with a single observation above it for a positive fluctuation), obtaining 2154 sources. The redshift distribution is shown in Fig.~3.
We used this sample to build a new Hubble Diagram for quasars up to z$\sim$6, which is shown in Fig.~4. 
The distance moduli have been calculated assuming the linear relation $\log(L_X)=\alpha\times\log(L_{UV})+\beta$, and using L=F$\times$4$\pi$$D_L^2$, where $D_L$ is the luminosity distance. From these relations, we easily obtain:
\begin{equation}
\log(D_L)=\frac{1}{2-2\alpha}\times(\log(F_X)-\alpha\times \log(F_{UV}))+\beta'
\end{equation}
where $\alpha$ is estimated as the average of the values obtained from the fits in small redshift intervals shown in Fig.~1, and $\beta'$ is a free parameter. If no absolute calibration is available, $\beta'$ cannot be determined, and we can only use the shape of the Hubble Diagram to constrain the cosmological parameters such as $\Omega_M$ and $\Omega_\Lambda$, with no information on $H_0$. Otherwise, we can fit the Hubble Diagram of quasars together with that of SN1A and use the overlapping redshift range to cross-calibrate the quasar diagram, so determining $\beta'$.
\begin{figure}
\includegraphics[width=83mm]{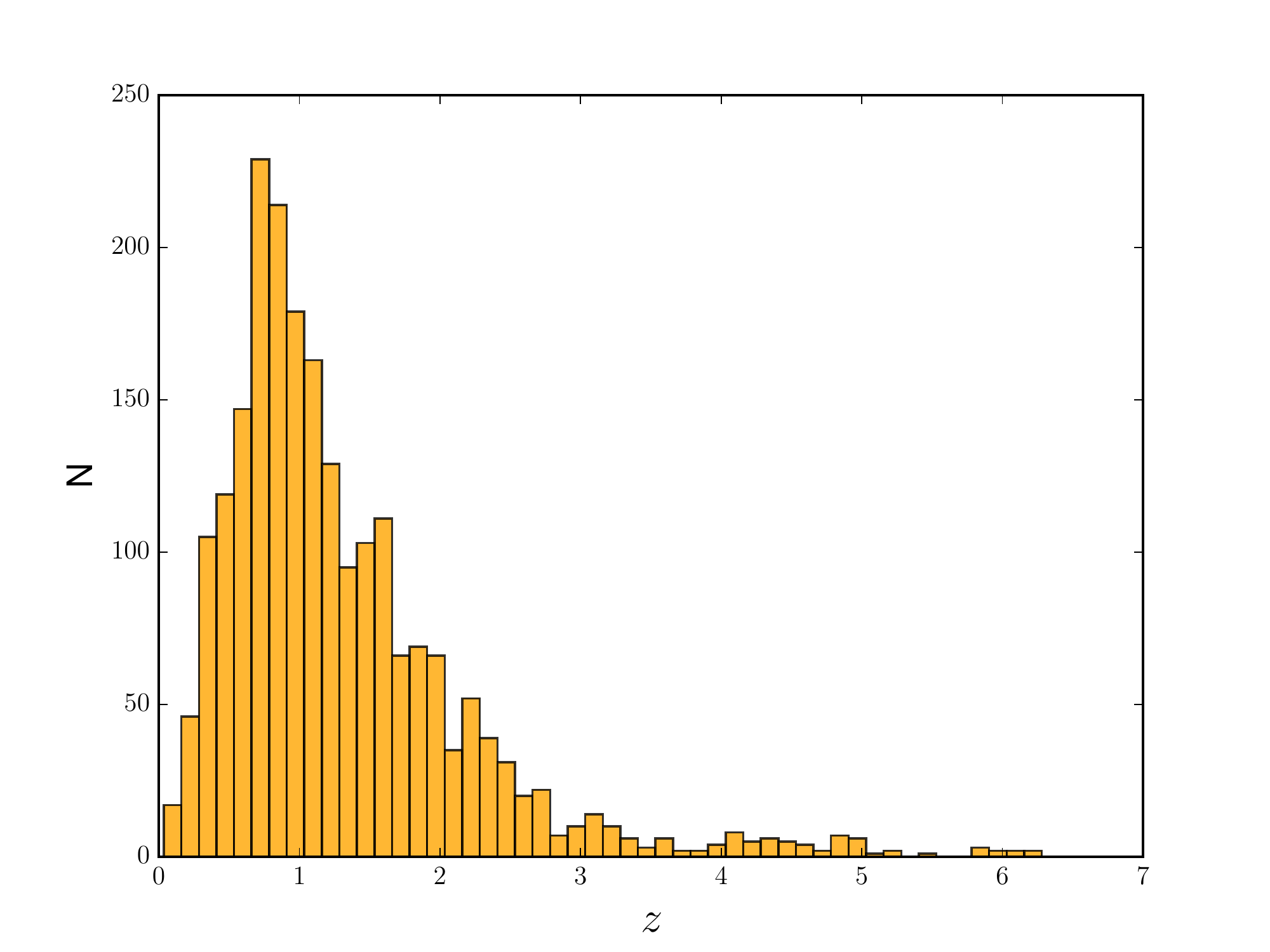}
\caption{Redshift distribution of the cleaned sample of 2177 quasars used to build the Hubble diagram of quasars.}
\label{label1}
\end{figure}
It is worth noticing that the ``quality filters'' mentioned above have been applied only up to z=3. For the few available quasars at higher redshift, no filter has been applied. This approach is justified by the little statistical weight of these objects in cosmological fits (see next Section), due to their small number ($\sim$50) and their high dispersion ($\sigma>$0.4). On the other hand, their presence  can be useful for a visual check of the Hubble Diagram at high redshifts.

\begin{figure}
\includegraphics[width=83mm]{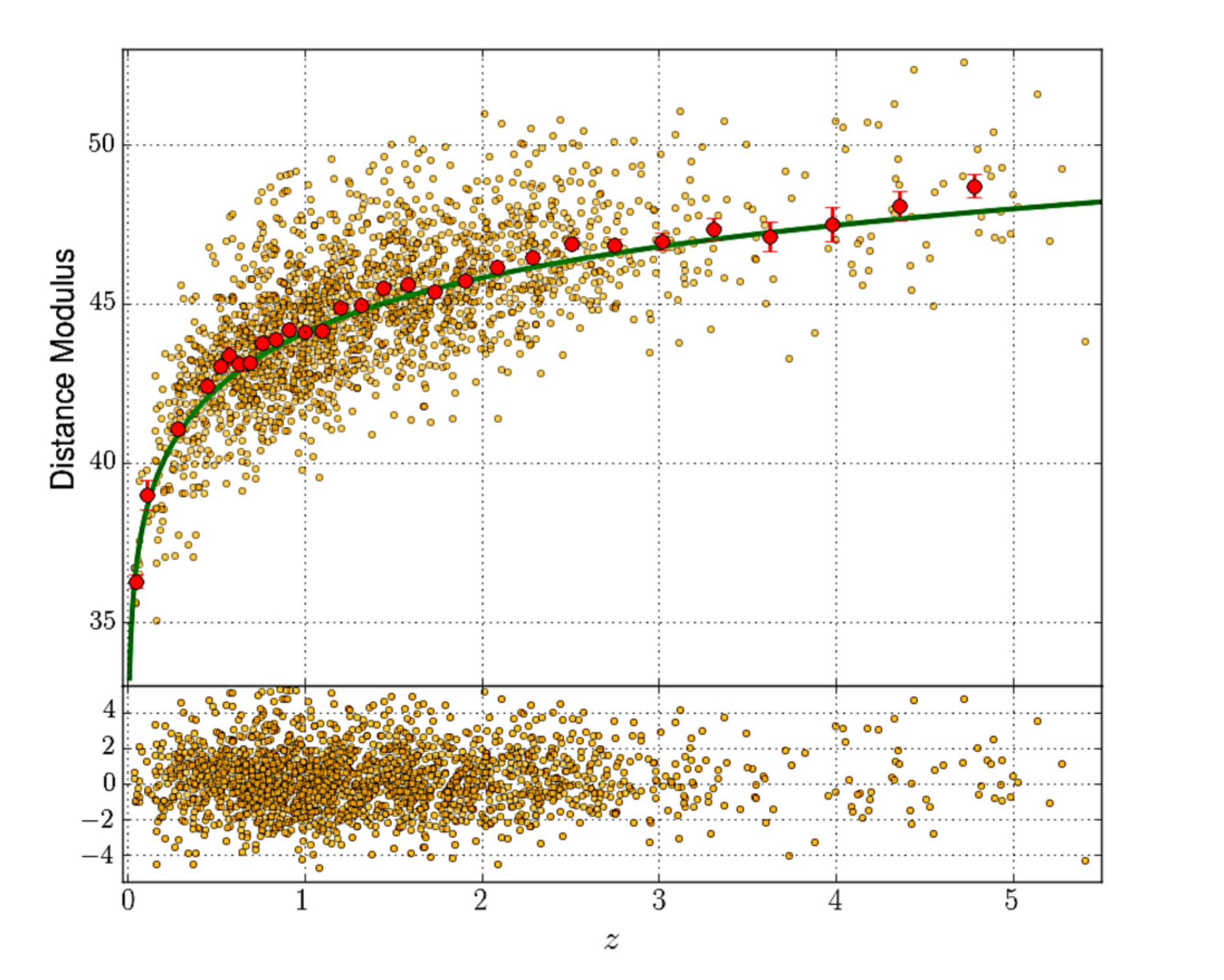}
\caption{Hubble Diagram for a ``cleaned'' sample of 2154 quasars. The gold points represent single quasars. The big red points are averages on small redshift intervals. }
\label{label1}
\end{figure}
\section{Cosmological estimates}
It is possible to test cosmological models, and measure cosmological parameters within a chosen model, by fitting the luminosity distance $D_L$ as a function of redshift. Alternatively, we can directly fit the F(X)-F(UV) relation:
\begin{equation}
\log(F_X)=\alpha\times\log(F_{UV})+\beta+(2-2\alpha)\log(D_L) 
\end{equation}
leaving the cosmological parameters and the parameters of the relation, $\alpha$ and $\beta$, as free parameters. In both methods, the fit has been performed adding an extra parameter $\delta$, representing the intrinsic dispersion of the relation (See Risaliti \& Lusso~2015 for details). Both methods produce fully consistent results.

The cosmological models fitted to the data are: (1) a flat $\Lambda$CDM model, (2) a non-flat $\Lambda$CDM; (3) a flat model with a constant dark energy equation of state w (w=-1 is equivalent to model 1); (4) a flat model with an evolving equation of state of dark energy: $w=w_0+w_a\times(1+a)$, where $a=(1+z)^{-1}$.

In all cases we fitted the quasar data together with SN1A from the JLA survey (Betoule et al.~2014), except for model 1, where quasars alone can reasonably constrain the value of $\Omega_M$. In joint quasar+SNe fits we allowed for a free cross-normalization parameter.
Our findings are summarized in Table~1.

\begin{table}
\caption{Results from fits to the Hubble diagram of SNe and quasars}
\label{tlab}
\begin{tabular}{lcccc}\hline
Model &$\Omega_M$&$\Omega_\Lambda$&w(or w$_0$)&$w_a$\\ 
\hline
flat $\Lambda$&0.28$^{+0.04}_{-0.05}$&&&\\
open $\Lambda$&0.25$^{+0.06}_{-0.06}$&0.67$^{+0.11}_{-0.11}$&&\\
w&0.25$_{-0.10}^{+0.08}$&&-0.93$^{+0.19}_{-0.25}$&\\
w$_0$w$_a$&0.30$^{+0.13}_{-0.09}$&&-0.91$^{+0.20}_{-0.26}$ & 0.8$^{+1.5}_{-2.7}$\\
\hline
\end{tabular}
\end{table}

The most relevant result of our analysis is shown in Fig.~5, where we plot the w$_0$-w$_a$ contour for the evolving dark energy model. Even if the constraints on the two parameters from quasars and SNe alone are not tight (Table~1), their combination with the constraints from the other main cosmological probes  (CMB, BAO, weak lensing, Planck collaboration~2015) improve (for the moment slightly) the errors on both w$_0$ and w$_a$. This is mainly due to the fact that quasars add complementary information with respect to the other methods, mainly related to the redshift interval z=1-3. This is the interval where the possible effect of a non-zero w$_a$ parameter are higher, and is not probed by any of the other methods. BAO constraints at z$>$2 from Ly$\alpha$ absorption is not included in the contours of Fig.~5, but they are likely to add important new information (Aubourg et al.~2015). 
\begin{figure*}[ht!]
\includegraphics[width=56mm]{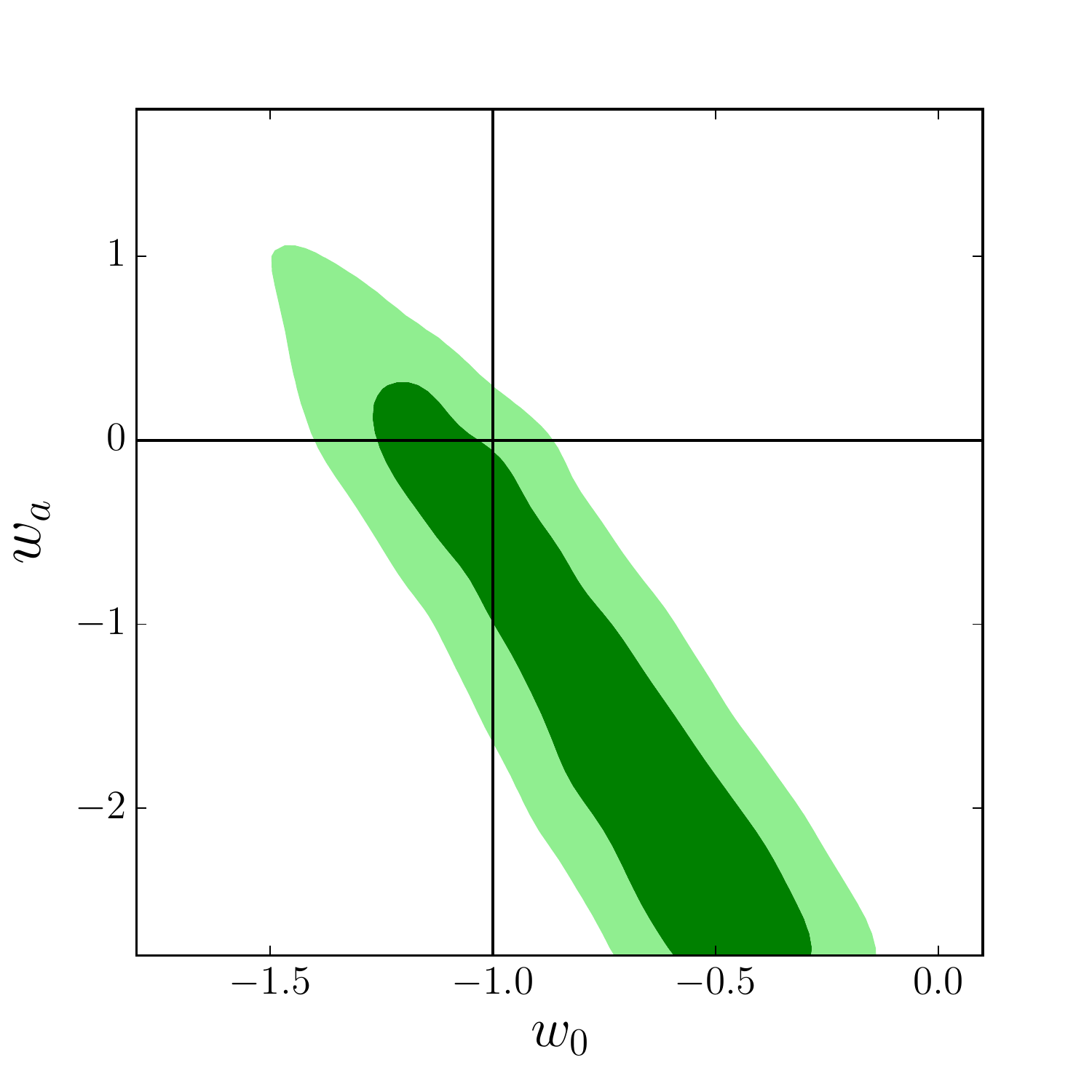}
\includegraphics[width=56mm]{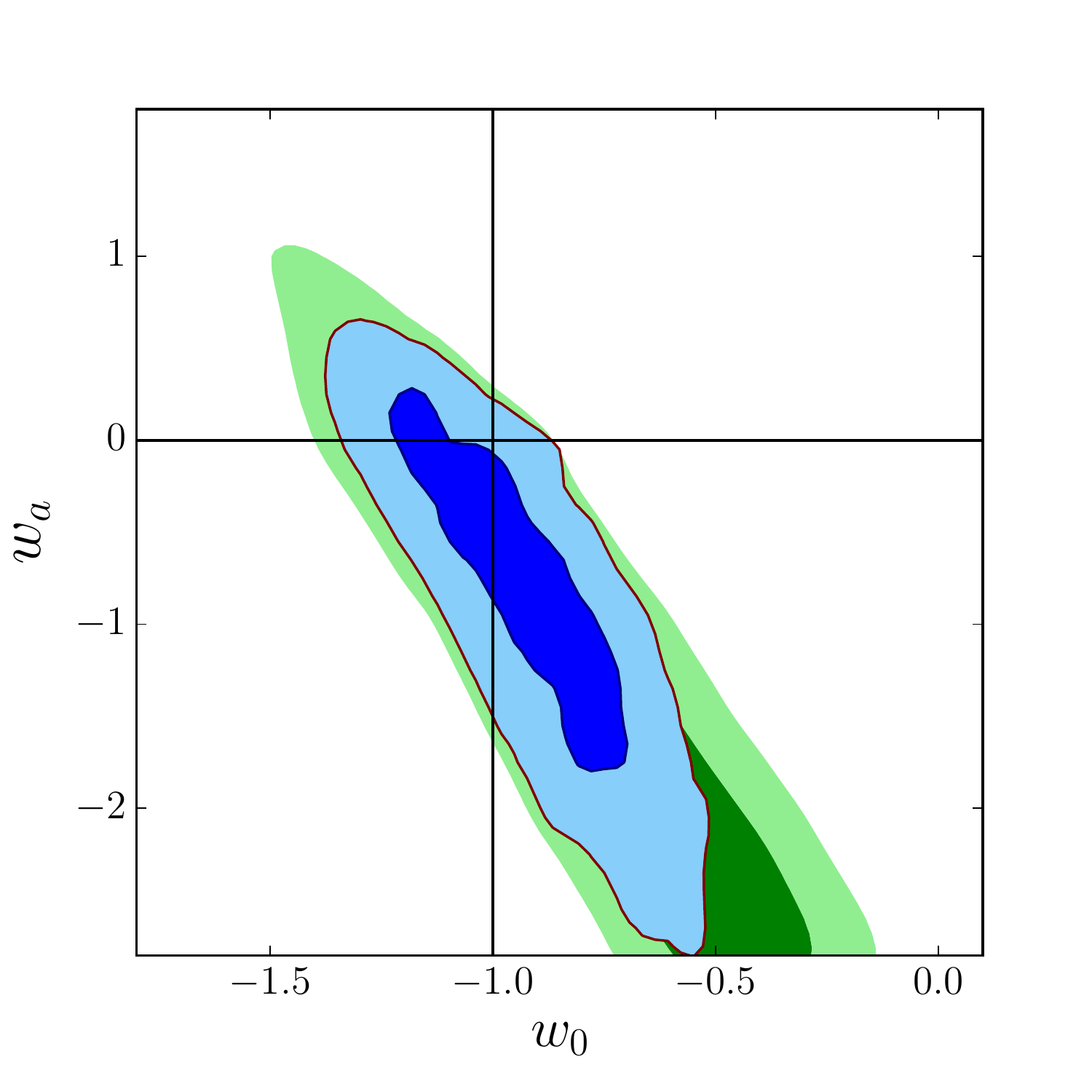}
\includegraphics[width=56mm]{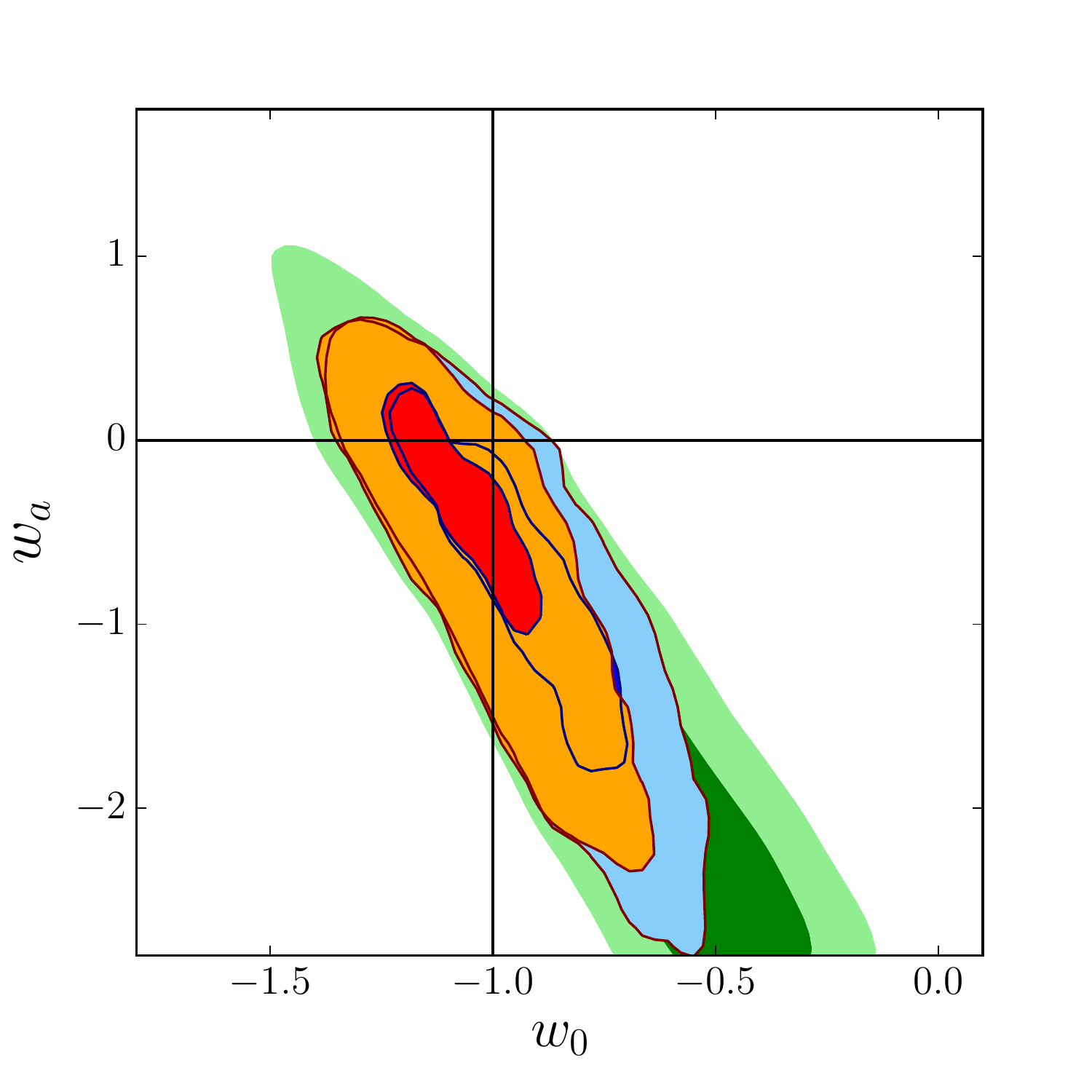}
\caption{Left: w$_0$-w$_a$ contours obtained from the combined Planck, BAO and JL constraints (Planck collaboration~2015); Middle: same, adding constraints from SNe (Betoule et al.~2014); Right: same, further adding the constraints from the Hubble diagram of quasars. }
\label{label1}
\end{figure*}

The limited improvement shown in Fig.~5 is an indication that with the currently available statistics quasars are just starting to be significant in constraining the properties of the dark energy component. However, our work is still quite incomplete. First, we can make our filters more efficient in order to recover more objects in our ``good'' sample: now the efficiency is of the order of 25\%, due to our choice to be extremely conservative in order to avoid systematic effects. We plan to better investigate the issue of systematic effects through simulated samples, and to improve the efficiency up to a factor of $\sim$2. Second, we will soon include several hundreds of new sources with {\em Chandra} and SDSS observations, by cross-correlation the {\em Chandra} Point Source Catalogue with the SDSS quasar samples.
Moreover, in the next few years large surveys such as the XXL (Pierre et al.~2016) will add a few thousands of new quasars, and the new X-ray survey mission eROSITA will provide hundreds of thousands of sources. 

\section{An XMM-Newton next decade program on high-z quasars}

The work presented in the previous Sections has been based on   serendipitous X-ray observations. This is certainly a winning strategy at low redshifts. However,   the low surface density and the low flux of high redshift quasars make them increasingly rare in serendipitous searches. The Hubble Diagram in Fig.~2 contains only $\sim$50 objects at z$>3$ (not filtered, as explained in the previous Section), and this number drops to a few units if the same quality filter as for z$<$3 objects is applied. 

On the other hand, nearly all-sky optical surveys like the SDSS provide thousands of quasars at z$>$3. These objects have a too low surface density to be effectively found in serendipitous X-ray observations (the whole XMM-Newton archive covers a few hundreds square degrees), but the brightest of them have an expected 2-10~keV flux of a few 10$^{-14}$~erg~cm$^{-2}$~s$^{-1}$, enough to obtain a good XMM spectrum with observations of a few tens of ks. This opens a new window of opportunity for future XMM-Newton observations: if in the next years a significant amount of observing time is devoted to z$>3$ quasars, a unique advance in testing cosmological models can be achieved. 

A key point to underline here is the importance of cosmological probes sensitive to a different combination of cosmological parameters with respect to the currently available methods. If the ``true'' cosmological model is the standard $\Lambda$CDM model, with a cosmological constant which is just an intrinsic geometrical property of space-time, we already have a very satisfactory measurement of $\Omega_M$ and $\Omega_\Lambda$. Therefore, the most useful new cosmological probes are those able to detect possible deviations from the standard model, through constraints on new regions of the parameter space. Filling the Hubble Diagram at redshift z$>$3 provides exactly this kind of constraints. Taken alone, it has a limited cosmological information (at z$>$3 the Universe is dominated by dark matter), but this information 
is related to the dark energy parameters in a different way than other cosmological probes. In terms of contour errors, it provides a contour with a major axis almost orthogonal to that of present measurements (Fig.~4). Therefore, even if the area of the new contour is large, the intersection with the currently available ones can reduce the uncertainty by a large factor.
This is illustrated by a simulation of 100 quasars with redshift in the interval 3.0-3.5, assuming the L$_X$-L$_{UV}$ relation and a cosmological model with evolving equation of state of dark energy and w$_0$=-0.8 and w$_a$=0.5. These values are allowed by current constraints (Fig.~5). When these new quasars are added to the Hubble diagram and used to fit the cosmological parameters, we obtain the results shown in Fig.~6, with the standard w$_0$=-1, w$_a$=0 model ruled out with a statistical significance of $\sim$3~$\sigma$, and the dark energy ``figure of merit'' (i.e. the inverse of area of the w$_0$-w$_a$ contour) increased by a factor 2.5 with respect to present constraints.
 \begin{figure}
\includegraphics[width=83mm]{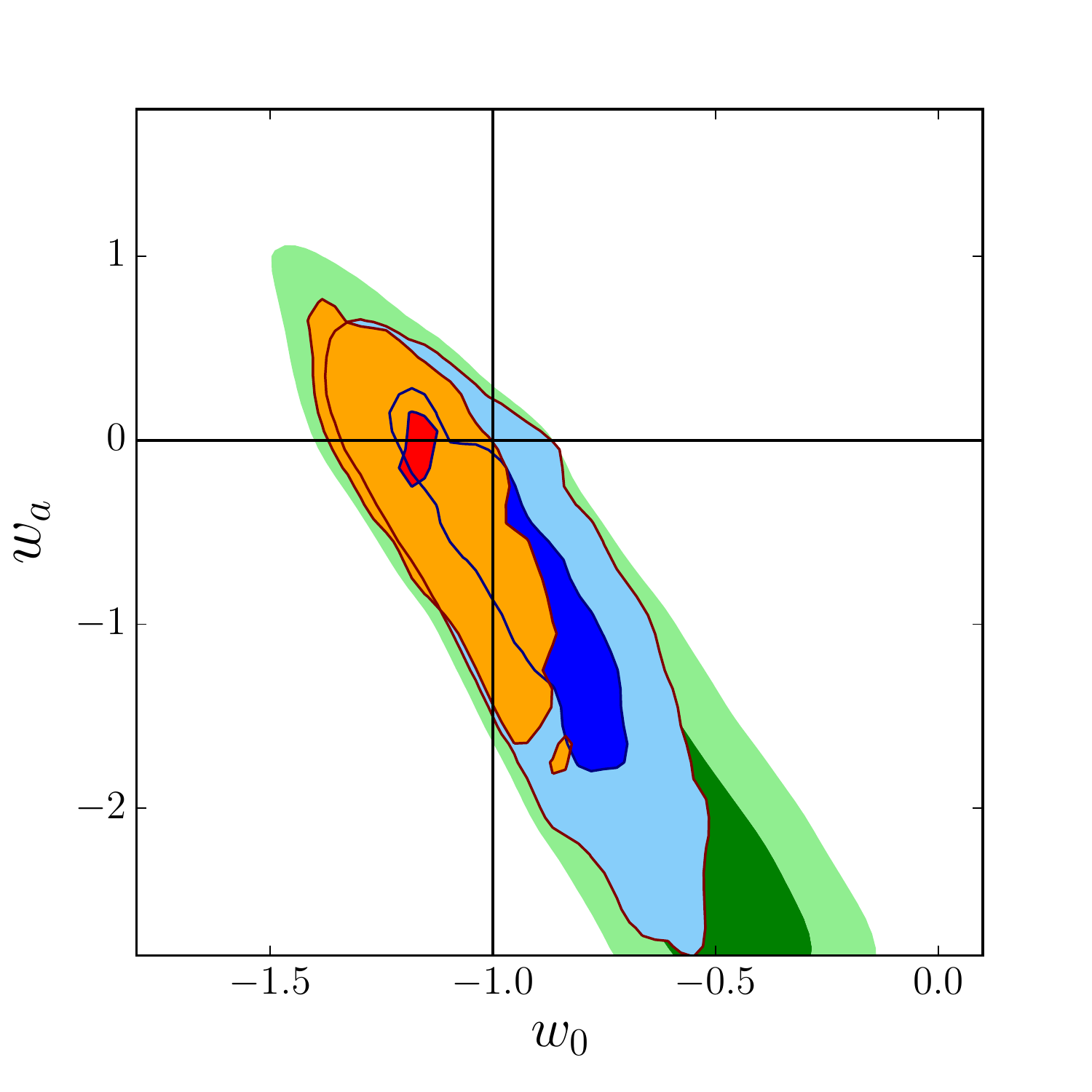}
\caption{w$_0$-w$_a$ contour (same as in Fig.~5c) obtained by adding to our current quasar sample a group of  100 simulated quasars with redshift z=3.0-3.5, assuming a w$_0$w$_a$CDM model with w$_0$=-1.2 and w$_a$=0.5.}
\label{label1}
\end{figure}

Besides cosmology, observations of bright, optically selected high-z quasars will be quite relevant for our understanding of the spectral properties of these sources. Indeed, the information on the X-ray emission of quasars at $z>3$ is still quite incomplete. The   Chandra-COSMOS legacy survey (which is arguably the highest quality sample of high-z quasar available, Marchesi et al.~2016), contains only 11 type~1 quasars at z$>$3 with a high enough S/N for a basic spectral analysis. Even in these cases, the spectra can provide only limited information on the properties of these objects, as deduced by the high uncertainty in the determination of the spectral index (Fig.~14 of Marchesi et al.~2014). Basically, there is no good spectrum of z$>$3 quasars. The pointed observations of optically selected quasars would drastically change this scenario, thanks to the higher brightness of the selected objects.  This would allow for the first time a comparison of  z$>$3, very luminous quasars with lower redshift and/or lower luminosity sources, providing an important test of possible evolutionary effects.

\section{Conclusions}
In this paper we have shown that the non-linear L$_X$-L$_{UV}$ relation in quasars (log[L$_X$]$\sim$0.6~log[L$_{UV}$]+8.5) has an observed intrinsic dispersion of about 0.2~dex. We can use the non-linearity of the relation to derive quasar distances, and to build a Hubble diagram of quasars up to z$\sim6$. Fitting a distance-redshift relation it is then possible to test cosmological models and measure cosmological parameters. We showed that the sample available today merging literature data and the cross-match between SDSS quasar catalogues and the XMM-Newton 3XMM-DR6 catalogue of serendipitous X-ray sources, is enough, if used together with supernovae, to constrain the cosmological parameters of a w$_0$w$_a$CDM model, slightly improving the current contours. 

We also showed that, while at low redshift (z$<$2.5) serendipitous searches are the most efficient way to build large quasar samples with X-ray and UV measurements, at z$>3$ pointed observations of UV-selected sources are the only way to obtain significant constraints on the distance-redshift relation. With the observation of $\sim$100 z$>3$ quasars we would significantly improve the current constraints on the dark energy equation of state parameters (Fig.~6). 

Based on this work, we conclude that the observation and spectral characterization of quasars at high redshift, and its application to cosmology, may become a new important part of the XMM-Newton program for the next decade. 

\acknowledgements
This work has been supported by  the grant ASI - INAF I/037/12/0.

\end{document}